\documentclass[aps,prb,twocolumn,groupedaddress]{revtex4-1}
\usepackage{graphicx,xcolor}
\usepackage{gensymb}
\usepackage{multirow}

\begin{document}

\title{First-principles structure search for the stable isomers of 
       stoichiometric WS$_2$ nano-clusters}

\author{Roohollah Hafizi}
\email{r.hafizi@ph.iut.ac.ir}

\author{S. Javad Hashemifar}
\email{hashemifar@cc.iut.ac.ir}

\author{Mojtaba Alaei}
\email{m.alaei@cc.iut.ac.ir}

\author{MohammadReza Jangrouei}
\email{m.jangroii@ph.iut.ac.ir}

\author{Hadi Akbarzadeh}
\email{akbarzad@cc.iut.ac.ir}
\affiliation{Isfahan University of Technology, 84156-83111 Isfahan, Iran}

\date{\today}
\newcommand{\etal}{{\em et al}}

\begin{abstract}
In this paper, we employ evolutionary algorithm along with the full-potential 
density functional theory (DFT) computations to perform a comprehensive search for 
the stable structures of stoichiometric (WS$_2$)$_n$ nano-clusters ($n=1-9$),
within three different exchange-correlation functionals.
Our results suggest that $n=3$, 5, 8 are possible candidates for the low temperature 
magic sizes of WS$_2$ nano-clusters while at temperatures above 600 Kelvin, 
$n=5$ and 7 exhibit higher relative stability among the studied systems.
The electronic properties and energy gap of the lowest energy isomers
were computed within several schemes, including semilocal PBE and BLYP functionals,
hybrid B3LYP functional, many body based DFT+GW approach, 
and time dependent DFT calculations.
Vibrational spectra of the lowest lying isomers, computed by the force constant method,
are used to address IR spectra and thermal free energy of the clusters.
Time dependent density functional calculation in real time domain is applied 
to determine the full absorption spectra and optical gap of the lowest energy 
isomers of the WS$_2$ nano-clusters.
\end{abstract}


\maketitle

\section{Introduction}

Discovery of the hollow carbon structures in the period of 1985-1991,
\cite{kroto1985c, iijima1991helical} 
and their fascinating properties along with promising applications is assumed 
to be one of the major landmarks of contemporary science world. 
Further investigations revealed that these fantastic structures are in fact feasible 
isomers of all layered structure materials which are highly anisotropic;
\cite{tenne1992polyhedral, feldman1995high}
First experimental evidence on the so-called Inorganic Fullerene-like (IF) structures
was reported on MoS$_2$ and WS$_2$, in 1992.\cite{tenne1992polyhedral}
Molybdenum disulfide and tungsten disulfide belong to the family of 
Transition Metal Dichalcogenides (TMDs) with the general formula 
of MX$_2$ (M=Mo, W, Nb and X=S, Se, Te), 
which share similar layered structure known as Molybdenite, in their bulk form.
 
Although, lubricant properties had been the main application of WS$_2$ until some years ago, 
new fascinating properties was the reason for this material to be subject of 
many recent scientific studies.
While in its bulk form, WS$_2$ exhibits an indirect semiconductor band gap 
of around 1.35\,eV, \cite{kuc2011influence} 
it has been shown that an individual mono-layer of WS$_2$ has a direct band gap of around 2\,eV,
\cite{gutierrez2012extraordinary} 
which evidences its potential for photoluminescence applications
\cite{hill2015observation, gutierrez2012extraordinary}. 
On the other hand, lack of inversion symmetry in these monolayers gives rise to 
non-trivial non-linear optical responses.\cite{janisch2014extraordinary, torres2016third}
Potential applications in hydrogen energy producing cycle,\cite{voiry2013enhanced}
lithium ion battery,\cite{silbernagel1975lithium} and flexible electronics
\cite{liu2014elastic} are further motivations for several experimental
and theoretical investigations on layered WS$_2$ structures.
\cite{mai2014exciton, hong2014ultrafast, PhysRevB.92.115443,
gao2015large, PhysRevB.91.075205}

In spite of several studies on the nano-layered structures of TMDs, 
the atomic nano-clusters of these systems are not well investigated. 
In most cases, people have considered that a nano-cluster is a piece 
of a layer of original material.
\cite{helveg2000atomic, yang2003adsorption, lauritsen2007size, li2007electronic}
The limited ab initio studies on TMD nano-clusters are mainly concerned with some 
pre-selected structures, ignoring a reliable search for the stable structure of these systems.
\cite{singh2005novel, li2007electronic, murugan2005atomic, murugan2007assembling}
To the best of our knowledge, there is only one manual structure search on 
the stable structures of MoS$_2$ and WS$_2$ nano-clusters.\cite{murugan2005atomic}

In the current work, by combination of evolutionary algorithm and first-principles calculations, 
we have done a wide systematic structure search on (WS$_2$)$_n$ clusters ($n=1-9$). 
In addition, magic clusters are identified by considering a number of criteria, 
including second-order energy difference, HOMO-LUMO gap and binding energy. 
The vibrational properties and IR spectra of the energetically most stable structures 
are also addressed and the calculated vibrational free energy 
is used to investigate finite temperature effects on 
the stability of the clusters.

\section{Computational Method}

Systematic crystal structure prediction was performed using an evolutionary algorithm
developed by Oganov \etal. and features local optimization, real-space representation, 
and flexible physically motivated variation operators.
\cite{lyakhov2013new, oganov2011evolutionary} 
This algorithm starts with a trial generation of structures consisting of 
random configuration of atoms and/or some favorable \textit{seed} structures.
This generation is then provided to an auxiliary code 
to calculate their minimized total energies.
The calculated minimized total energies are used to filter out the unlikely structures 
and then make a new generation of structures by applying the heredity, mutation, 
permutation, and random operators.
This procedure is repeated until the most stable structures are found. 

The total energy calculations and minimizations were performed in the framework of 
Kohn-Sham density functional theory (DFT) by using the all-electron full-potential method 
implemented in FHI-aims package.\cite{blum2009ab, ren2012resolution}
This code employs numeric atom-centered orbital (NAO) basis functions
to achieve computational efficiency and numerical accuracy, 
especially for non-periodic systems. 
In order to increase the computational performance, 
the geometrical relaxation of the clusters was done in three steps: 
initial relaxation with no relativistic effects and a light basis, 
secondary relaxation by turning on the scalar relativistic effects 
and using a tight basis, 
and final relaxation with spin-polarized calculations.
Our comprehensive structure search were preformed within 
two generalized gradient approximations (GGA);
Becke-Lee-Yang-Parr (BLYP)\cite{lee1988development} 
and Perdew-Burke-Ernzerhof (PBE),\cite{perdew1996generalized}
and one hybrid functional; B3LYP.\cite{becke1993density}
The geometry relaxations were done by using the Broyden-Fletcher-Goldfarb-Shanno 
(BFGS) algorithm down to the residual atomic forces of less than 10$^{-2}$eV/\AA. 
The vibrational frequencies and IR spectra of the most stable isomers were calculated
by applying a finite displacement of about 10$^{-3}$\AA\ to atomic positions and 
using the force constants method. 

In order to obtain more accurate fundamental energy gaps, 
the many body based GW correction
was applied to the Kohn-Sham electronic structure of the most stable isomers.
In the GW approximation, for better description of many body correlations,
the Kohn-Sham quasi-particles are allowed to weakly interact via 
a screened coulomb potential.
In order to manage the heavy GW calculations, a smaller basis set (tier1) 
was used for this part of the computations.
Some benchmark calculations on smaller clusters, revealed that this choice of
basis set, compared with tier2, induces a typical error of about 0.2-0.3\,eV 
in the calculated energy gaps.
The optical absorption spectra and optical gaps of the stable isomers 
were calculated in the framework of time dependent DFT
by using the Octopus package.\cite{octopus:2003}
In this code, a real space grid and pseudopotential technique are
used for real time propagation of the Kohn-Sham orbitals.

\begin{table}[h]
\caption{
  Obtained energy order of (WS$_2$)$_n$ isomers within BLYP, PBE, and 
  B3LYP functionals. For any size ($n$), the first row shows the lowest energy isomer
  and other rows stand for meta-stable isomers.
  The energy difference (eV) of the meta-stable isomers with 
  the corresponding lowest energy isomer is written in the parenthesis.
}
\label{tab:isomer}
\begin{ruledtabular}
\begin{tabular}{llll}
 $n$  &   BLYP   &    PBE     &   B3LYP   \\
\hline
 2  & 2a         & 2a         & 2a         \\
    & 2b (0.196) & 2c ($<$meV)& 2b (0.229) \\\vspace{0.2cm}
    & 2c (0.341) & 2b (0.186) & 2c (0.503) \\
 3  & 3a         & 3a         & 3a         \\              
    & 3b (0.275) & 3c (0.624) & 3b (0.393) \\\vspace{0.2cm}
    & 3c (0.290) & 3b (0.705) & 3c (0.437) \\
 4  & 4a         & 4b         & 4a         \\              
    & 4b (0.057) & 4d (0.257) & 4b (0.082) \\
    & 4c (0.453) & 4a (0.294) & 4d (0.502) \\\vspace{0.2cm}
    & 4d (0.643) & 4c (0.585) & 4c (0.631) \\
 5  & 5a         & 5a         & 5a         \\              
    & 5b (0.059) & 5b (0.367) & 5b (0.275) \\
    & 5c (0.344) & 5c (0.642) & 5c (0.453) \\\vspace{0.2cm}
    & 5d (0.655) & 5d (0.727) & 5d (0.669) \\
 6  & 6a         & 6e         & 6a         \\              
    & 6b (0.316) & 6d (0.086) & 6c (0.470) \\
    & 6c (0.564) & 6c (0.338) & 6b (0.485) \\
    & 6d (0.691) & 6a (0.342) & 6d (0.619) \\\vspace{0.2cm}
    & 6e (1.501) & 6b (0.354) & 6e (1.456) \\
 7  & 7a         & 7a         & 7a         \\              
    & 7b (0.398) & 7d (0.116) & 7c (0.326) \\
    & 7c (0.469) & 7e (0.137) & 7b (0.563) \\
    & 7d (0.970) & 7c (0.182) & 7d (0.921) \\\vspace{0.2cm}
    & 7e (1.073) & 7b (0.670) & 7e (0.963) \\
 8  & 8a         & 8a         & 8a         \\              
    & 8b (0.343) & 8b (0.779) & 8b (0.416) \\
    & 8c (0.909) & 8d (1.247) & 8c (1.091) \\
    & 8d (1.066) & 8c (1.423) & 8d (1.171) \\\vspace{0.2cm}
    & 8e (1.285) & 8e (1.662) & 8e (1.617) \\
 9  & 9a         & 9e         & 9d         \\              
    & 9b (0.068) & 9f (0.437) & 9a (0.166) \\
    & 9c (0.138) & 9b (0.460) & 9e (0.214) \\
    & 9d (0.274) & 9a (0.460) & 9b (0.293) \\
    & 9e (0.410) & 9c (0.721) & 9c (0.374) \\
\end{tabular}
\end{ruledtabular}
\end{table}

\section{Results and Discussions}

\begin{figure*}
\includegraphics{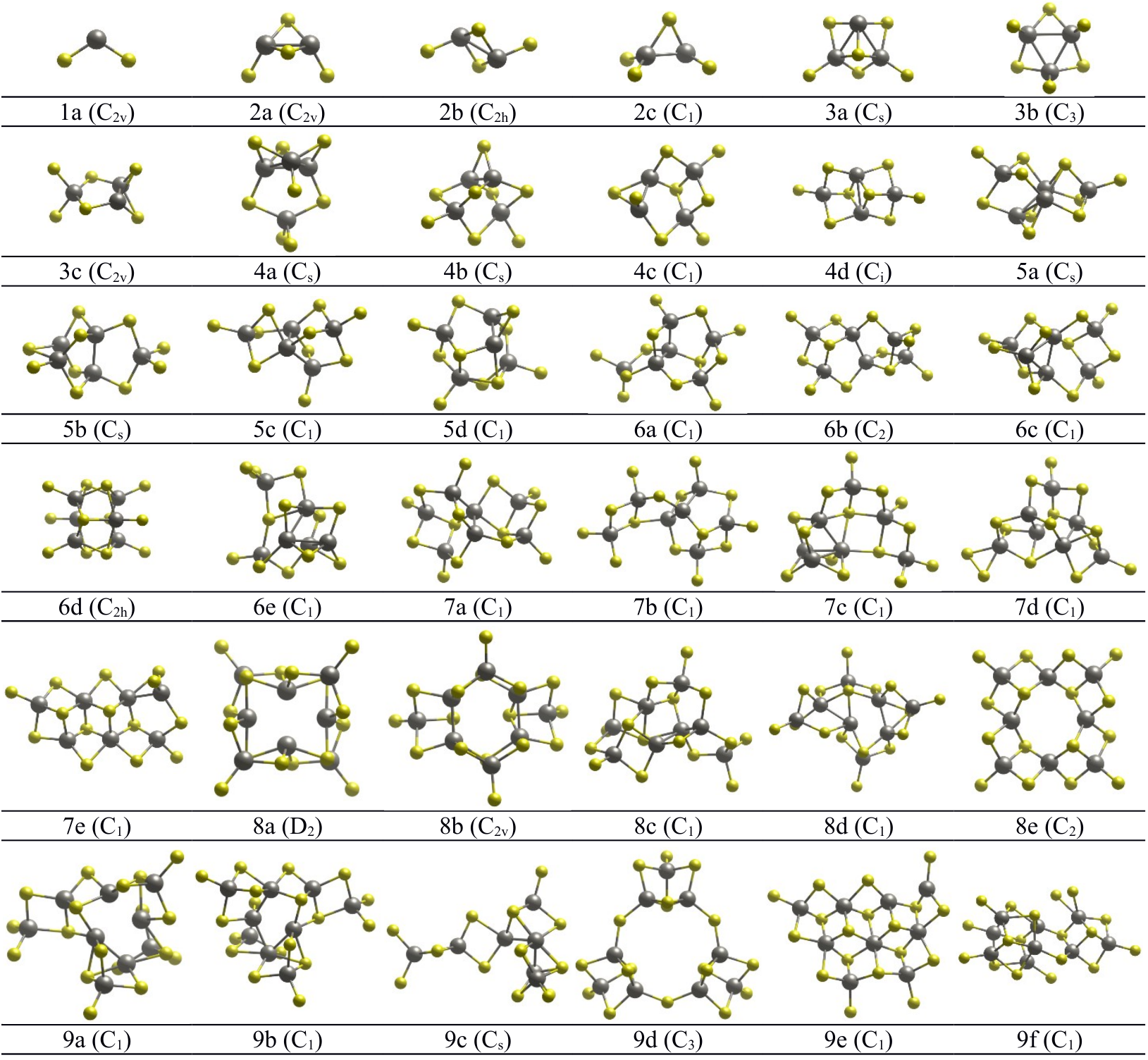}
\caption{\label{figure:isomer}
  Obtained lowest energy structures of (WS$_2$)$_n$ nano-clusters ($n=1-9$).
  The small light balls stand for S atoms while
  the larger dark balls show W atoms.
  The point group of the clusters, mentioned in parenthesis, is determined by using 
  MacMolPlt software.\cite{bode1998macmolplt}
}
\end{figure*}

The more stable structural isomers of (WS$_2$)$_n$ clusters are shown in 
Fig.~\ref{figure:isomer},
sorted by their minimized energy within the BLYP functional.
The energy order of the obtained isomers within PBE and B3LYP is
compared with BLYP in table~\ref{tab:isomer}.
The energy difference of the isomers with 
the most stable isomer is also presented in the table.
We observe that B3LYP generally follows BLYP and predicts very 
similar set of lowest energy isomers.
Except (WS$_2$)$_9$, the lowest energy isomer of other nano-clusters is 
the same within BLYP and B3LYP,
and the second isomer of BLYP usually coincides with that of B3LYP.
On the other hand, PBE exhibits more differences with BLYP;
the lowest energy isomer of 4th, 6th, and 9th clusters within PBE is different with BLYP.
Moreover, the second isomer of the most clusters within PBE is also 
different with BLYP. 
The following discussion, unless explicitly mentioned, 
is mainly based on the BLYP calculations.

\begin{figure}
\includegraphics[scale=0.48]{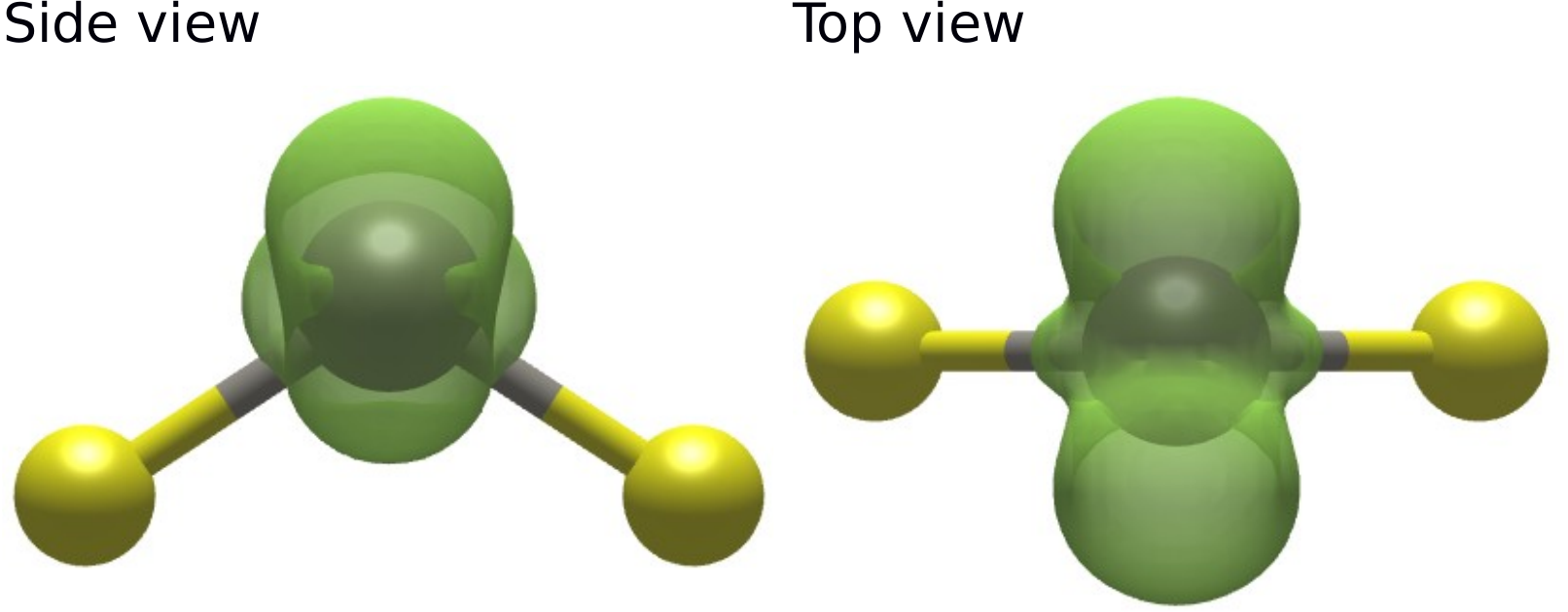}
\caption{\label{figure:spin}
  Calculated spin density of WS$_2$, 
  green areas show the high spin density regions.}
\end{figure}

WS$_2$ molecule ($n=1$) prefers an obtuse bond angle of about $108.4\degree$ which 
is attributed to the lone pairs of electrons in the outer electronic
shell of tungsten atom.
The W-S bond length in WS$_2$ is 2.11\,\AA\ which is approximately 12\% shorter than 
the bulk value (2.41\,\AA\ for 2H-WS$_2$),\cite{schutte1987crystal} 
which is due to the lower coordination of atoms in the cluster.
In the bulk WS$_2$, each tungsten atom bonds to six sulfur atoms, hence individual
W-S bonds in the bulk system, compared with WS$_2$ cluster, is weaker and longer.
We found that BLYP and B3LYP predicts a second isomer for WS$_2$ (not shown) which has 
an acute angle of about $55\degree$ and a minimized energy that is about 2.91\,eV
higher than the ground state isomer. 
In addition, we found that although a linear S-W-S isomer is symmetrically stable, 
it has large unstable vibrational modes.
The lowest energy structure of WS$_2$ exhibits an spin polarized electronic structure
with a net spin moment of 2\,$\mu_B$, mainly originated from the W $d$ electrons. 
The calculated spin density of this system (Fig. \ref{figure:spin}) shows that
spin polarization is distributed around the tungsten atom, perpendicular
to the bond direction.
Hence, spin polarization comes from the less hybridized part of the W $d$ electrons.
We found that all other (WS$_2$)$_n$ clusters ($n>1$) prefer a non-magnetic ground state,
which is likely due to the higher coordination and consequently 
more hybridization of tungsten atoms in these systems.
Murugan \etal. have reported magnetization in their lowest isomer of (WS$_2$)$_6$.
We checked their magnetic cluster within different XC functionals and 
found that this cluster has an energy of 1.27, 0.57, and 3.97\,eV higher 
than our lowest energy isomer of (WS$_2$)$_6$
within BLYP, PBE, and B3LYP, respectively.

In the size of $n=2$, the most stable isomer (2a) is the one with two sulfur atoms 
at the bridge positions and two others at the terminus positions (Fig.~\ref{figure:isomer}).
Within PBE, the 2c isomer with one bridge and three terminus positions 
has almost the same energy as 2a and is more stable than 2b (table~\ref{tab:isomer}),
while within BLYP the energy of 2c is about 0.34\,eV 
higher than 2a (table \ref{tab:isomer}). 
For (WS$_2$)$_3$ cluster, the structure with a sulfur atom at the cap position has 
the lowest energy, while more symmetric 3b isomer has a higher energy of about 0.27\,eV. 
It is again seen that 3c is more stable than 3b within PBE. 
It seems that structures with more planar spatial extension 
are more favorable within PBE, compared to BLYP and B3LYP.
The lowest energy structure of (WS$_2$)$_3$ disagrees with a previous 
theoretical study by Murugan and others.\cite{murugan2005atomic}
In the case of larger clusters also, a significant difference is observed between our
lowest energy isomers and those of Murugan \etal.,
their lowest energy isomers are usually more symmetric.
It is speculated that their manual structure search has been limited to 
the more symmetric geometries 
and hence they have missed some more stable low symmetry configurations. 
For example, we found that their most stable isomer of (WS$_2$)$_6$ 
relaxes to a 6d like structure, which is a distorted version of the original structure.
In other words, the original structure prefers a one dimensional elongation.
We attribute this kind of distortions to the Jahn-Teller effect,\cite{englman1972jahn} 
which reduces symmetry of the systems and enhances their stability.

Because of the layered structure of bulk WS$_2$, layered isomers are expected 
to appear more frequently as the cluster size increases and formation of 
core-shell clusters is expected to be unfavorable.
It is seen that 7e, 8b, 8e, 9d, and 9e isomers exhibit planner layered 
configurations; all W atoms in these clusters lie on the same plane.
This observation confirms the considerable tendency of WS$_2$ nano-clusters 
to form 2D configurations, even in small sizes. 
The forth isomer of (WS$_2$)$_6$ (6d, second isomer within PBE) 
displays an interesting ring shape structure; 
the coordination number of all W atoms is 6 (as in the bulk).
 
It is noticed that several low energy isomers of 
our bigger clusters (some are not shown in Fig.~\ref{figure:isomer})
display \textit{semi-planar} tilde like structures. 
For accurate identification of the semi-planar structures, we compared the RMS error 
of a second order polynomial fit to the cluster atomic positions 
with that of a WS$_2$ mono-layer.
The 8a and 8b isomers are examples of these semi-planar structures. 
Some preliminary calculations indicate that structural deviation
of these systems from planner geometry may be due to 
the unsaturated W bonds and hence adding S atoms to these clusters
enhances formation of planner configurations.
The investigation of non-stoichiometric WS$_2$ clusters
is a separate ongoing project which will be presented in the future.

Since there is no reference experimental data on WS$_2$ nano-clusters, 
the choice of a proper XC functional can be a major question. 
Hybrid functionals yield somewhat smaller deviations from experiment 
for 4d and 5d element compounds,\cite{bühl2008geometries,waller2007geometries} 
however, they are demanding and impractical for big clusters. 
Comparing the obtained results within BLYP, PBE, and B3LYP reveals 
a good agreement between BLYP and B3LYP results. 
Thus, although PBE functional reproduces the electronic properties
of WS$_2$ mono-layer correctly,\cite{gutierrez2012extraordinary}
it seems that BLYP is a proper less demanding functional for 
first-principles study of WS$_2$ nano-clusters.

\begin{figure}
\includegraphics*[scale=0.90]{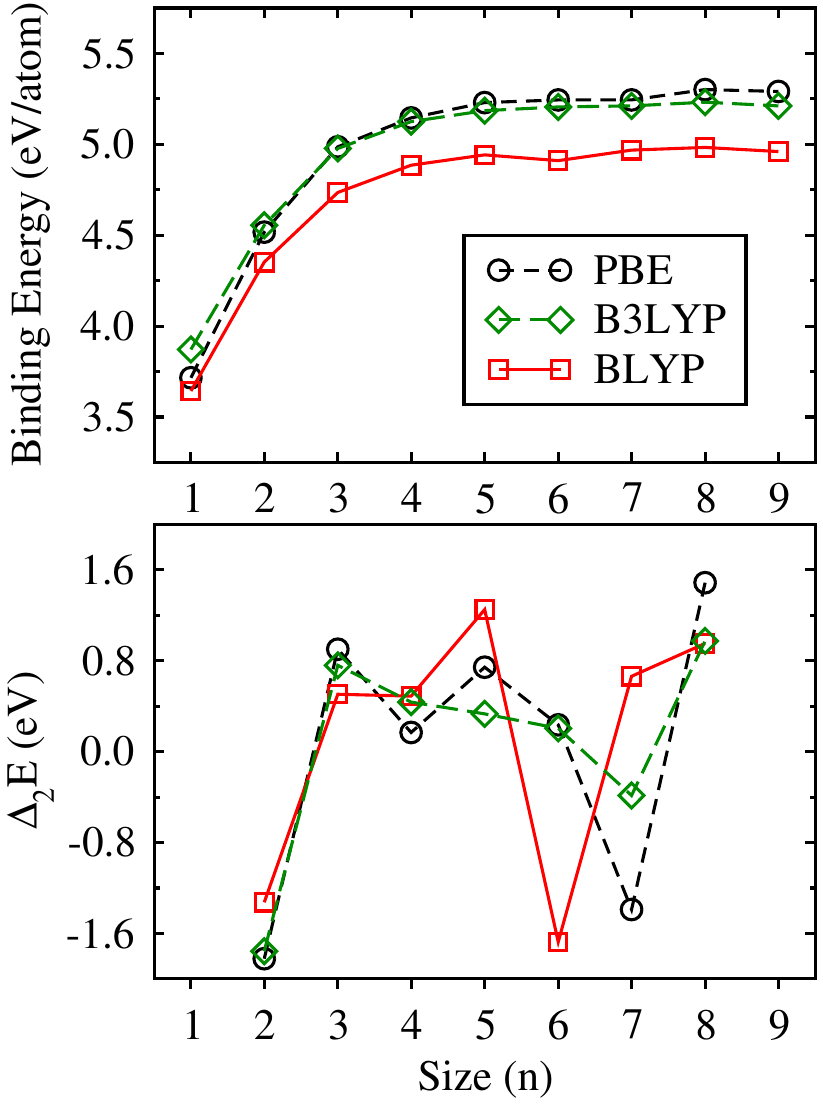}
\caption{\label{figure:energy}
  Binding energy and second order energy difference ($\Delta_2 E$) 
  of the lowest lying isomers of (WS$_2$)$_n$ clusters 
  within different XC functionals.}
\end{figure}

In order to address stability of the clusters, the absolute binding energy per atom
of their lowest energy isomer was calculated and presented in Fig.~\ref{figure:energy}.
Overall behavior of absolute binding energy (BE) is a monotonic increment with size,
indicating more stability of larger clusters.
It is attributed to the fact that increasing the size decreases
relative number of the surface broken bonds of the cluster. 
Taking into account the tendency of the lowest lying isomers to have 
a semi-planar configuration, 
it seems that BE is logically converging to the BE of a WS$_2$ mono-layer.
There is a local minimum at $n=6$ in BLYP, which indicates lower relative
stability of (WS$_2$)$_6$, compared with its neighboring clusters. 
The same reasoning is applicable to (WS$_2$)$_7$ within PBE.

A conventional parameter to describe the relative stability of clusters is
the second-order difference of energy: 
$\Delta_2 E = E_{n+1} +E_{n-1}-2E_{n}$ , $E_n$ being 
the minimized energy of the cluster of size $n$.
From a physical point of view, $E_{n+1}+E_{n-1}-2E_{n}$ measures 
the tendency of two cluster of size $n$ to form two clusters of sizes $n+1$ and $n-1$; 
taking into account the absolute energy values, 
a positive value of $\Delta_2 E$ indicates higher relative stability
of the corresponding cluster with respect to the neighboring sizes.
The calculated $\Delta_2 E$ values for WS$_2$ clusters 
are shown in Fig.~\ref{figure:energy}.
It is seen that the most stable sizes or the so called magic sizes are 
$n=5, 8$ within BLYP, $n=3, 8$ within B3LYP, 
and $n=3,5,8$ within PBE calculations.
Our results does not confirm the previous reported magic size of 6
in WS$_2$ clusters.\citep{murugan2005atomic} 
It is noteworthy to state that these results belong to the zero Kelvin temperature, 
and we will investigate finite temperature effects, after discussing
vibrational properties of the lowest lying isomers.

\begin{figure}
\includegraphics[scale=0.9]{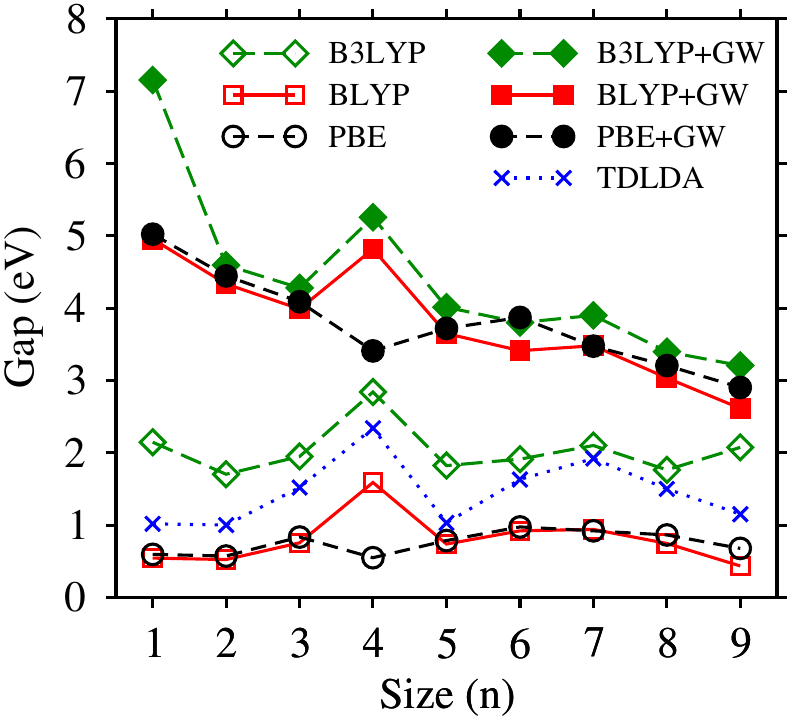}
\caption{\label{figure:gap}
  Calculated HOMO-LUMO gap of the lowest lying isomers within 
  different XC functionals, before and after application of
  the GW correction.}
\end{figure}

The calculated HOMO-LUMO gaps of the lowest lying isomers within the three functionals
used in this paper, before and after application of the many body based
GW correction, are presented in Fig.~\ref{figure:gap}.
It is seen that the main impact of the GW correction is an upward shift of the gap,
which is more pronounced in the smaller clusters.
Before GW correction, the overall trends within BLYP and B3LYP are very similar,
while the absolute values within BLYP coincides with the PBE data (except for $n=4$).
Application of the GW correction is seen to decrease the difference between
the three functionals.
We found that (WS$_2$)$_4$ occurs on a local maximum of the HOMO-LUMO gap
diagram within the BLYP and B3LYP functionals, before and after the GW correction.
It is an indication for the higher chemical hardness of this cluster,
compared with the other sizes.
Hence, in spite of the small $\Delta_2 E$ value of this cluster,
we speculate that (WS$_2$)$_4$ should be visible in the envisaged mass spectra
measurements on WS$_2$ nano-clusters.
The GW corrected gaps exhibit an overall decrease toward the band gap 
of a WS$_2$ mono-layer (~2\,eV).
This trend supports our earlier statement that WS$_2$ cluster geometries 
approaches the sandwich structure of the bulk WS$_2$ sheets, as the size increases.

\begin{figure}
\includegraphics*[scale=0.90]{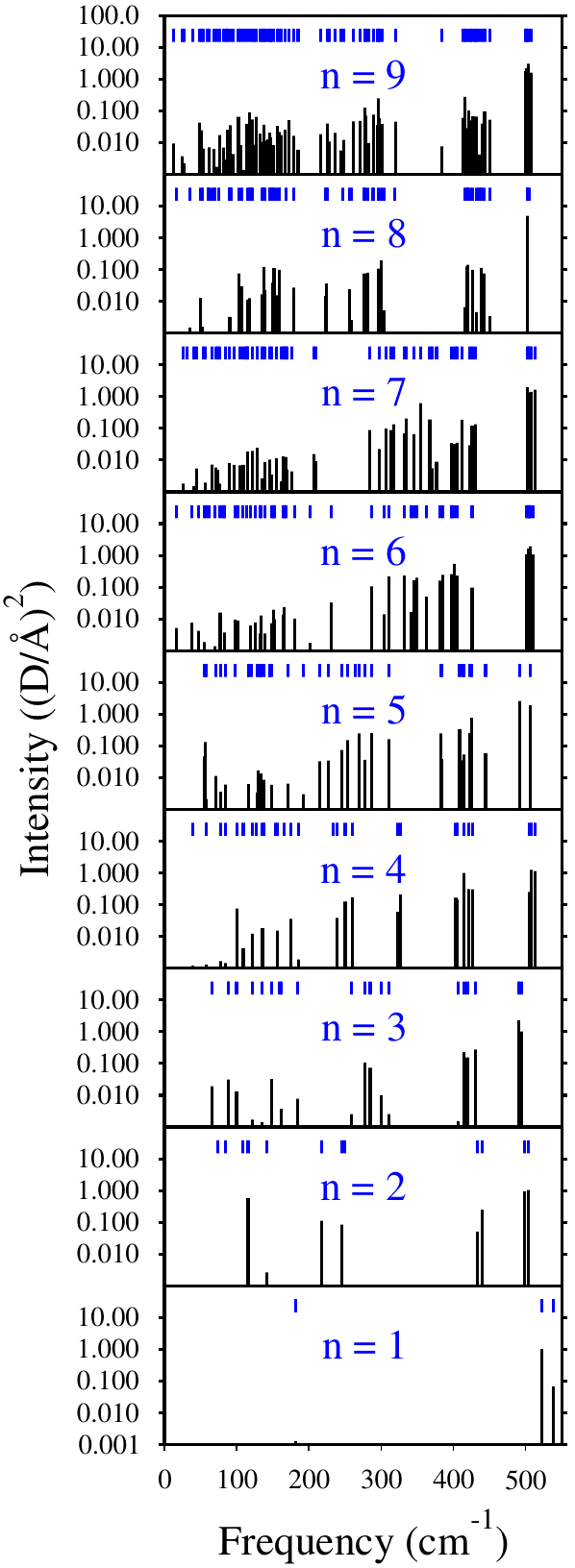}
\includegraphics*[scale=0.66]{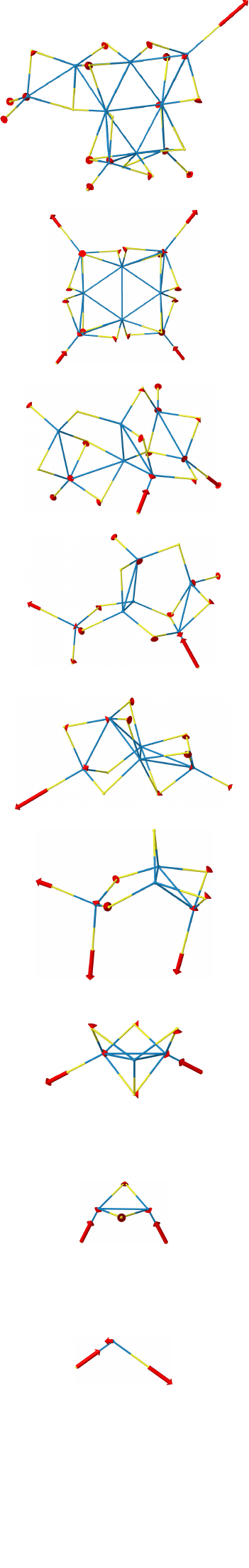}
\caption{\label{figure:vib}
  Calculated IR spectra of the lowest energy isomers of (WS$_2$)$_n$ clusters
  within BLYP. Please note the logarithmic scale of the Intensity axis.
  The blue ticks inside upper border of the plots show the vibrational modes of 
  the corresponding system.
  In the right side, the dominant IR mode of the clusters is sketched. 
  Red arrows show instantaneous movement direction of atoms with 
  an amplitude proportional to the length of arrow.
}
\end{figure}

The dynamical properties of the lowest lying isomers were investigated by
calculating the vibrational modes of the systems and identifying their IR spectra,
in the framework of force constant method.
In this method, a finite displacement of 10$^{-3}$\,\AA\ was applied to
all atomic positions of the fully relaxed structures and
the obtained Hessian matrices were diagonalized to obtain
the vibrational modes of the clusters.
The corresponding IR intensities, given in Fig.~\ref{figure:vib},
were calculated by derivation of the dipole moments along the vibrational modes.
Appearance of no negative (imaginary) frequency in the obtained vibrational spectra
implies dynamical stability of the lowest energy isomers.
Schematic atomic oscillations of the lowest lying isomers in their dominant IR frequency
are presented in Fig.~\ref{figure:vib}.
The dominant IR frequency takes a value around 500\,cm$^{-1}$ in all sizes,
and is almost one of the hardest mode of the clusters.
We found that the dominant modes (Fig.~\ref{figure:vib})
are mainly consisted of some dangled W-S bonds oscillation.
Since tungsten is about six times heavier than sulfur,
oscillation amplitude of sulfur atom is much larger than tungsten atom.
In a dangled W-S bond, a higher degree of hybridization happens between 
tungsten and sulfur electrons and hence dangled bonds are expected to be 
stronger than other bonds.
As a result of that, these bonds are the origin of the hard vibrational 
modes of the clusters.

\begin{figure}
\includegraphics{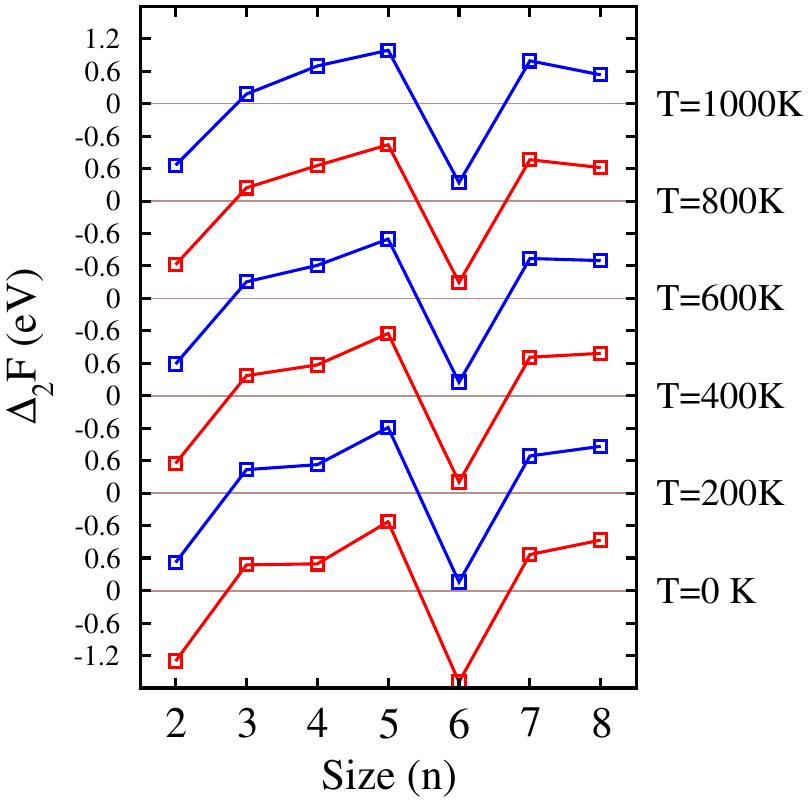}
\caption{\label{figure:d2f}
  Calculated vibrational free energy of the stable isomers 
  at different temperatures ranging from 0 to 1000\,K.}
\end{figure}

\begin{figure}
\includegraphics[scale=0.9]{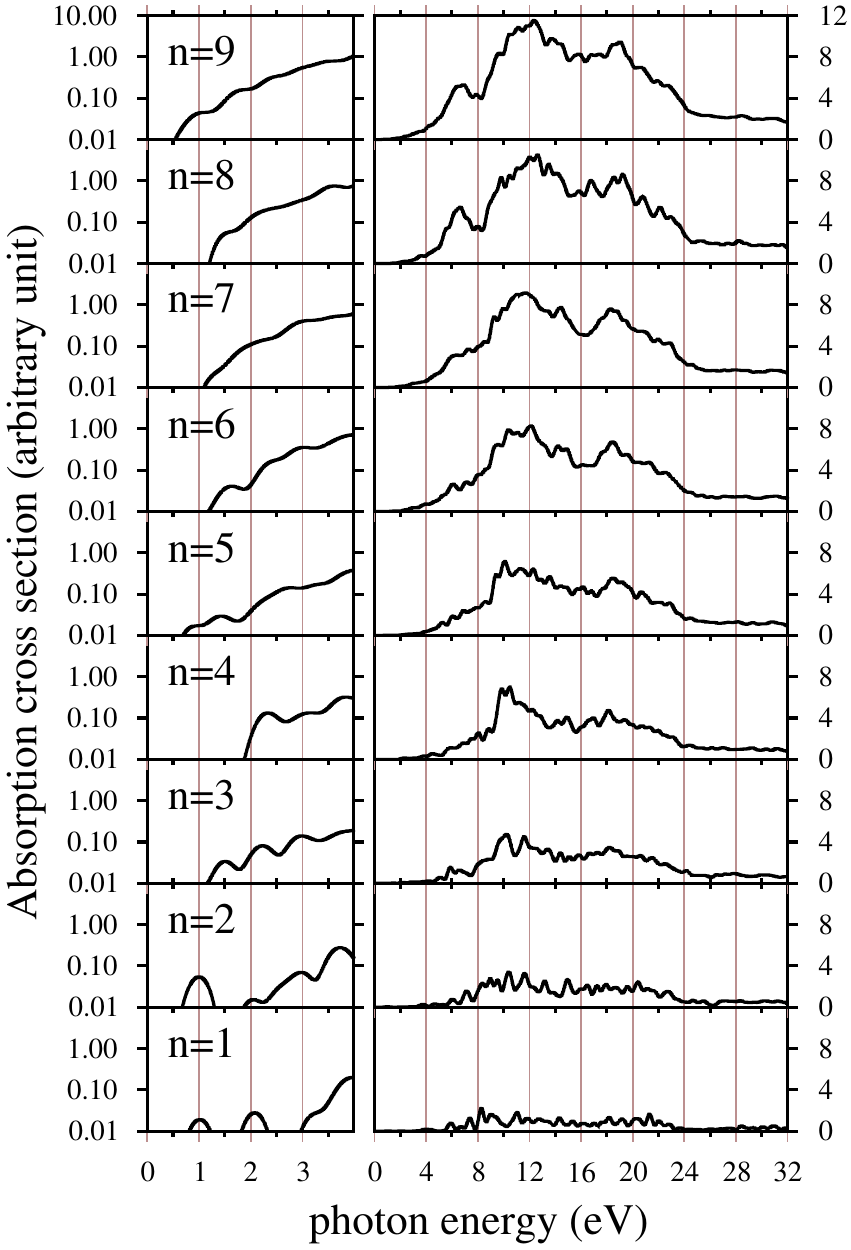}
\caption{\label{figure:optic}
  Calculated optical absorption cross section of the stable isomers of
  WS$_2$ nano-clusters. The full spectrum is presented in the right side
  while for accurate determination of the optical gap,
  initial part of the spectrum with a logarithmic scale 
  is presented in the left side.}
\end{figure}

In order to investigate the influence of thermal vibrational excitations on 
relative stability of the clusters at elevated temperatures, 
we calculated the vibrational Helmholtz free energy (F) 
of the systems from their obtained vibrational spectra, as follows:
\begin{center}
$F=E+ \frac{1}{2}\sum\limits_{i}^{3N}\varepsilon_i+k_BT \sum\limits_{i}^{3N}\ln(1-e^{-\beta\varepsilon_i})$
\end{center}
where $E$ is the total energy of the system, $i$ runs over the number of vibrational modes, 
$\varepsilon_i$ is the $i$th mode energy, $N$ is the number of atoms in the cluster, 
$k_B$ is the Boltzmann constant, $T$ is the Kelvin temperature, and $\beta$ equals $1/k_BT$. 
The Helmholtz free energy was calculated in different temperatures,
ranging from zero to 1000\,K. 
Then, the second order difference of the Helmholtz free energy ($\Delta_2F$) 
was determined at different temperatures (Fig.~\ref{figure:d2f}).
It is noticed that thermal effects enhance relative stability of (WS$_2$)$_4$ 
and (WS$_2$)$_7$ clusters and decrease relative stability of (WS$_2$)$_3$ cluster.
The reason is that (WS$_2$)$_4$ and (WS$_2$)$_7$ have more low energy vibrational modes,
compared with their neighboring clusters (Fig.~\ref{figure:vib}).
As a result of that, vibrational entropy of these systems increases more rapidly
by increasing temperature.  
Consequently, increasing temperature enhances relative stability of these two nano-clusters.
At temperatures above 600\,K, (WS$_2$)$_7$ becomes an important magic system
while (WS$_2$)$_3$ seems not to be a magic cluster anymore.
We see that the main magic cluster, (WS$_2$)$_5$, keeps its high relative stability
up to temperatures of about 1000\,K.

The optical absorption spectra of the lowest energy isomers of WS$_2$ clusters
were calculated by using time dependent Kohn-Sham approach
in the real time domain.
The obtained absorption cross sections are presented in Fig.~\ref{figure:optic}.
We observe that by increasing the size, two broad peaks appear at 
energies of about 12 and 18\,eV and one rather sharp peak appears
at energy of about 6\,eV.
Following our statement that the clusters are converging to a WS$_2$ sheet,
these three peaks may be attributed to a periodic WS$_2$ monolayer.
In order to determine the optical gap of the clusters, 
we tried to identify the first allowed optical transition of the systems.
In this regard, the initial part of the absorption spectra is
enlarged and plotted in a logarithmic scale in Fig.~\ref{figure:optic}.
The predicted optical gap of the systems are presented in Fig.~\ref{figure:gap}.
It is seen that the optical gaps follow well the trend of the HOMO-LUMO gaps 
within BLYP while the optical gap values are somewhere between
the corresponding BLYP and B3LYP values.
Moreover, (WS$_2$)$_4$ and (WS$_2$)$_7$ are found to be on the local
maxima of the optical gap diagram, indicating high relative chemical
stability of these two clusters.

\section{Conclusions}

In this paper, we searched for the best structures of (WS$_2$)$_n$ nano-clusters 
(n = 1 - 9) using evolutionary algorithm.
The total energy calculations and structural relaxations were performed
in the framework of density functional theory (DFT), 
using full potential numeric atom centered orbital technique.
The calculations were performed within two semilocal (PBE and BLYP) and 
one hybrid (B3LYP) functionals.
It is discussed that BLYP is likely the proper functional for 
first-principles investigation of the WS$_2$ nano-clusters.
We found that beyond $n = 3$, the clusters prefer extended 2D sandwich like structures,
and formation of core-shell configurations is ruled out. 
Except for WS$_2$, no spin polarization was seen in the lowest lying isomers of the other clusters. 
The obtained vibrational free energies within BLYP, indicate that the zero temperature 
magic sizes of WS$_2$ clusters are $n=5$ and 8, 
while it was argued that at finite temperatures, high vibrational entropy of (WS$_2$)$_7$ 
enhances relative stability of this system to become a magic cluster above 600\,K.
The magic cluster (WS$_2$)$_5$ was found to keep its high relative stability 
up to 1000\,K.
The many body based GW correction was applied to improve the obtained energy gap of 
the clusters within the selected semilocal and hybrid functionals.
It was found that (WS$_2$)$_4$ exhibits a high chemical hardness, 
among the studied systems.
The obtained absorption cross sections, in the framework of time dependent DFT,
indicates that the optical gap of the clusters follows well the trend
of the energy gaps within the BLYP functional.

\begin{acknowledgments}
This work was jointly supported by the Vice Chancellor
of Isfahan University of Technology (IUT) in Research Affairs and
ICTP Affiliated Centre in IUT.
\end{acknowledgments}

\bibliography{references}

\end{document}